\documentstyle[preprint,prl,aps]{revtex}

\begin{document}
\draft
\title{Mean Field Calculations of Bose-Einstein Condensation
of $^7$Li Atoms in a Harmonic Trap}
\author{T. Bergeman}
\address{Physics Department, SUNY, Stony Brook, NY 11794-3800}
\date{\today }
\maketitle

\begin{abstract}
A self-consistent mean-field theory for bosons for $T>0$ is used to
reconcile predictions of collapse with recent observations of Bose-Einstein
condensation of $^7$Li.  Eigenfunctions of a (non-separable) Hamiltonian that
includes the anisotropic external trap field and atom-atom interactions are
obtained by an iteration process. A sum over the Bose distribution, and
the ``alternating direction implicit'' algorithm are used.  Near $T_{c}$,
the ensemble exhibits a localized condensate composed of atoms in the few
lowest states. For lower $T$, numerical instability indicates collapse to a
more dense phase.
\end{abstract}


Recent observations of Bose-Einstein condensation in tenuous gases of alkali
atoms \cite{Cornell,Hulet,Ketterle} have brought to concrete realization
a text-book paradigm and a prediction by Einstein 70 years ago.
These outstanding experimental achievements, coming after the rapid
development of methods for cooling atoms by laser light and by evaporation
in a trap, make it possible to test theories for weakly-interacting Bose
gases.  In contrast with superfluid helium, where the atoms interact strongly
and thus depart from simple predictions, there is hope that BEC in dilute
alkali atom gases can be usefully addressed by accessible theoretical
methods.

Perhaps the experimental result that presents the greatest challenge to
understanding is the reported Bose-Einstein condensation of $^7$Li, which
is known to have a scattering length, $a = - 27.2 a_{0}$ \cite{Abraham},
indicating an attractive atom-atom interaction.  It has been shown that in
free space, BEC with $a<0$ is not possible \cite{Stoof}, because of the
tendency to form a more dense phase.  For a harmonic trap, recent
computational results for $T=0$ indicate that a stable condensate can be
formed but with no more than about 1,400 atoms for the conditions of the
experiment \cite{Ruprecht,Baym,Stringari,Shuryak}.  When the number of atoms
exceeds this value, it is expected that the cloud will collapse into a
liquid or solid or into diatomic molecules. Since condensation of $^7$Li
has been reported with as many as 20,000 atoms \cite{Hulet}, there has been
speculation on various mechanisms that might account for the observations
\cite{Stringari,Shuryak,Greene,LYou}.

In this report, we present a self-consistent mean field
theory for trapped bosonic atoms.  Whereas many previous theoretical
studies concentrated on the condensate at $T=0$, we report fully quantal
results for $T>0$, even for $T>T_{c}$. From these results, we can study the
rapid changes in size and especially in the shape of the atomic cloud as the
temperature goes through $T_{c}$ for comparison with experimental
observations.  We will show that mean-field theory calculations with $^7$Li
parameters are consistent with the observed spatial distribution of cold
atoms in the trap \cite{Hulet}.

The Fermi-Thomas method or local density approximation (LDA) has yielded
many useful results for BEC \cite{Baym,Oliva,Chou}. This approximation
is an appropriate and efficient method for large numbers of atoms and
$a>0$.  On the other hand, the phenomena of interest with $a <0$ occur on
a scale equal to or less than the dimension of the lowest quantum state, so
the applicability of the LDA approach is severely limited in this case.
Other approaches to BEC for $^7$Li have involved three-body collisions
\cite{Greene} and higher order terms in the atom-atom interaction
constructed with derivatives of $\delta(\vec{r}_{12})$ \cite{LYou}.

The mean field approach to BEC is based on the pseudopotential form of
the atom-atom interaction \cite{HuangYang,Huang}.  We conjecture that
the atom-atom interaction for densities of interest ($a n^{1/3} \ll 1$)
can be written $V(\vec{r}_{12}) = (4 \pi a \hbar^2/M) \delta(\vec{r}_{12})$
even for $a<0$. From the Li$_{2}$ energies \cite{Abraham,Cote}, the next
term in the scattering series, the scattering range, may be computed.  We
find that it becomes significant at mK temperatures, but our concern here
will be restricted to $T < 1 \mu$K.

The discussion of the present computational approach can begin with the
Gross-Pitaevskii equation (GPE) for $T=0$ \cite{Gross,Pitaevskii}. With the
above form for the atom-atom interaction, the GPE for $N$ atoms of mass
$M$ is
\begin{equation}
\left[ -\frac{\hbar^2}{2M} \tilde{\nabla}^2 + V(\vec{\tilde{ r}}) +
\frac{4 \pi \hbar^2 a}{M} N |\psi(\vec{\tilde{r}})|^{2}\right]
\psi(\vec{\tilde{r}})  = \tilde{\mu} \psi(\vec{\tilde{r}}).
\end{equation}
Here $V(\vec{\tilde{r}})$ is the external harmonic potential and
$\tilde{\mu}$ the chemical potential.  $\psi$ is normalized so that
$\int d\vec{\tilde{\tau}}|\psi|^2 = 1$.
For applications to the trap used for $^7$Li atoms, the external potential,
$V = (M/2)(\omega_{z}^{2}\tilde{z}^2 + \omega_{\rho}^2 \tilde{\rho}^{2}) =
(M \omega^2 /2)(\lambda_{z}^2 \tilde{z}^2 + \lambda_{\rho}^2 \tilde{\rho}^2)$,
where $\omega = (\omega_{z} + 2 \omega_{\rho})/3$ is the average oscillator
frequency and $\lambda_{i} = \omega_{i}/\omega$ for $i = z$ or $\rho$.  For
a convenient distance scale, we let $\vec{\tilde{r}} = \alpha
\vec{r}$, with $\alpha^2 = \hbar/(2M \omega)$ and then let $\mu =
\tilde{\mu}/\hbar \omega$.  With $\int d\tau |\phi(\vec{r})|^2 = 1$, and
since $d\tilde{\tau} = \alpha^{3}d\tau$, $\phi(\vec{r}) = \alpha^{3/2}
\psi(\vec{\tilde{r}})$.  Thus the scaled GPE equation is
\begin{equation}
 \left[ -\nabla^2 + \frac{1}{4}( \lambda_{z}^2 z^2 +
\lambda_{\rho}^2 \rho^2 ) +  g N |\phi(\vec{r})|^2 \right] \phi(\vec{r})
= \mu \phi(\vec{r})
\end{equation}
The coefficient is $g = 8 \pi a/\alpha = 8 \pi a \sqrt{2 M \omega/\hbar}$.
$^7$Li is notable not simply because $a<0$, but also because $|g|$ is
relatively small.  For $^{23}$Na, $|g|$ is about 6 times, and for $^{87}$Rb,
$|g|$ is about 14 times the value for $^{7}$Li.  So in this respect,
$^7$Li is significantly closer to an ideal gas than the heavier
alkali atoms.

For $T>0$, one needs to obtain the wavefunctions for many orbitals
self-consistently, analogous to the Hartree-Fock theory of atoms.  Basic
equations for this approach can be found in Ref. \cite{FetterW}.  With
the pseudopotential approximation, the Hamiltonian equation for each basis
state in a self-consistent set is $H\phi_{i}(\vec{r}) = \epsilon_{i}
\phi_{i}(\vec{r})$, with
\begin{equation}
H  = - \nabla^2  + \frac{1}{4} \lambda_{z}^2 z^2 + \frac{1}{4}
\lambda_{\rho}^2 \rho^2 + g \sum_{j} N_{j} |\phi_{j}(\vec{r})|^{2}
\end{equation}
In the sum over states, the Bose distribution law applies:
\begin{equation}
N_{i} = \frac{g_{i}}{\{\exp{[(\epsilon_{i}
-\epsilon_0)\hbar \omega -\mu) /kT]}-1\}}
\end{equation}
where $g_{i}$ is the degeneracy and $\epsilon_{i}$ the energy eigenvalue
of the $i$th state.  The chemical potential,  $\mu$, is determined from the
condition $\sum_{i} N_{i} = N$.

The distribution law, Eq. (4), is often derived from the grand canonical
distribution based on an exchange of particles and energy between a large
ensemble and a subensemble. This situation seems irrelevant to a gas
of trapped atoms, which (after the evaporative cooling is turned off)
ideally exchanges neither particles nor energy with the surroundings.
We have therefore considered a microcanonical ensemble defined to include
all configurations of harmonic oscillator quantum states consistent with a
given (small range of) total energy and atomic number.  The number of
such possibilities increases very rapidly, so that with only 5 atoms and
$T$ = 20 nK for the trap of Ref.[2], more than 10$^{7}$ configurations
were found.  The distribution obtained strongly resembled the Bose
distribution with slight differences not likely to affect the present
discussion.  Since it is difficult to extend this microcanonical
distribution analytically to thousands of atoms, we have adopted the Bose
distribution. However, this point warrants further study.

Solutions to (3) are obtained by two stages of iteration.  For a chosen
$T$ and $\mu$, it is first assumed that $\phi_{i}(\vec{r})$ is given by
harmonic oscillator eigenfunctions for each set of quantum numbers
$N_{x}(i), N_{y}(i), N_{z}(i)$. The mean field is computed from the sum.
Since the Schr\"{o}dinger equation for $H$ (Eq. (3)) is nonseparable,
the alternating direction implicit (ADI) method is used with successive
scans in the $x, y$ and $z$ directions \cite{Lapidus}.  ADI iterations have
the effect of transforming a function $\phi_{x}(x)\phi_{y}(y)\phi_{z}(z)$
into a non-factorizable eigenfunction $\phi(x,y,z)$ of $H$. With
cylindrical coordinates, factors of $\rho^{m + 1/2}$ make numerical
differention near $\rho = 0$ unreliable, so Cartesian coordinates are used.
For $a<0$, the atom-atom interactions produce a potential well that is
typically smaller in spatial extent than the lowest orbital, so only the
lowest quantum states are significantly affected by the term in $g$.
Therefore, we needed to apply ADI to only the 4 to 50 lowest states. Other
states are retained in the sum in the form of the harmonic oscillator (HO)
functions.  Typically states up to $E = 6 kT$ are needed, so as many as
10$^7$ HO states were included.

In comparing calculations with experimental observations, it is pertinent
to recall features of Bose distributions that differ from the more familiar
Boltzmann forms.  For a given $T$, there is a range
of possible $\mu$ values, and the spatial distribution of the ensemble will
depend on $\mu$ as well as on $T$.  This is illustrated in Fig. 1, which shows
the distribution over $\rho$ calculated for a harmonic trap with the
parameters of the permanent magnet trap used for $^7$Li.  The frequency is
is 112 Hz in the axial direction and 149 Hz in the transverse
direction [2].  In this figure, the lighter curves are for $a$=0 (an ideal
gas), while for the heavier curves, $a = -27.2 a_{0}$, as for $^7$Li.  The
attractive interaction slightly narrows the distribution at large $N$.
In the limit $-\mu \rightarrow \infty$, one has effectively a Boltzmann
distribution.  As $\mu$ approaches 0, $N$ increases and the distribution
shrinks due to the onset of condensation. Thus from Fig. 1, it is apparent that
$T$ cannot be accurately deduced from the size of an atom cloud in a trap
unless $N$ is also known.  A similar statement applies to measurements of the
width of the velocity distribution, since the transformed wavefunctions
$\bar{\phi}_{i}(\vec{p})$ could have been used in place of $\phi_{i}(\vec{r})$
in the summation over discrete states.

Formation of a condensate with a fixed number of atoms as $T$ is varied
is shown in Fig. 2, for 20,000 atoms and $a$ = -27.2 $a_{0}$.  The condensate
peak develops over a very narrow range of $T$. As $T$ is reduced to slightly
below 162 nK, the computation becomes numerically unstable due to the
attractive atom-atom interactions. Similarly, for 1D calculations at $T=0$
\cite{Ruprecht,Baym,Stringari,Shuryak}, no eigenfunction can be found
beyond a certain critical value of $gN$.  The maximum number of atoms in the
ground state with the anisotropic trap and with the ADI algorithm is about
1,000, or somewhat less than obtained with an isotropic
trap of the same mean harmonic frequency.

Because of this instability, the calculations indicate that for more than
about 1,000 atoms, phase-coherent BEC in a trap cannot be achieved
with $^7$Li.  However, the term ``Bose-Einstein condensation'' often refers
to phenomena occurring over the entire range $T \leq T_{c}$ rather than
simply at $T=0$ \cite{Huang}.  In a harmonic trap, the signature of BEC
is the rapid decrease of cloud size or narrowing of the velocity
distribution, and these phenomena occur as the temperature is lowered through
$T_{c}$.  A general view of this phenomenon is shown in Fig. 3.
The filled circles on this figure represent $\rho(1/e)$ values taken from
the distributions shown in the previous figure (calculated
with $a$=-27.2 $a_{0}$), while all the curves are for $a$=0.  The temperature
at which the transition occurs is little affected by an $a$ of this magnitude.
For $a=0$, $\rho(1/e)$ drops to a limiting value equal to the size of the
ground state, but for $a=-27.2 a_{0}$, the collapse has no limit.
Determination of the ultimate minimum $T$ for a given $N$ would require an
investigation of fluctuation effects, which are beyond the scope of this
study.

The long dashed line in Fig. 3 shows that the sharp decrease in $\rho(1/e)$
occurs while $N_{0}/N < 10^{-2}$.  Thus it does appear that with $^7$Li, there
is a regime of $N$ or $T$ in which a ``condensate peak'' might be observable
but in which there is no threat of collapse into a more dense phase because
the accumulation of atoms in the ground state is far short of the critical
number.  From Fig. 3, the most obvious choice for a critical temperature
($T_{c}$) is 161 nK, which marks the limit of numerically stable solutions
for $a$ = -27.2 $a_{0}$. This is close to the value $T_{c} = 167$ nK from
semiclassical theory \cite{Bagnato} with the present trap parameters and $N$.

A consideration of Fig. 1-3 might suggest that $\rho(1/e)$ as a
measure of cloud size accentuates the condensate. If instead one plots the
average energy, $\langle E \rangle$, results for $a$=0 are shown in Fig. 3
as the short dashed line and inner left scale.  (We have divided by $3k$ for
easy comparison with $T$.) As discussed by deGroot et al.\cite{deGroot}, when
there is a finite
number of atoms in the ground state, $\langle E \rangle < 3kT$. In Fig. 3 there
is a discontinuity in the slope of $\langle E \rangle$ vs. $T$ at $T_{c}$,
so for $a$=0, the most significant decrease of $\langle E \rangle/3k$ occurs
below $T_{c}$.  $\langle E \rangle/3k$ does not mimic the sharp variation
in $\rho(1/e)$ because the condensate peak involves relatively few
atoms.

In the experiments \cite{Hulet}, the evolution of a fixed
ensemble as temperature is lowered is not observed.  Rather, atoms are
repeatedly placed in the trap and evaporatively cooled.  After each load
and cool cycle, the total number of atoms, $N$, is measured by optical
absorption, and also the size of the cloud is measured by imaging a probe beam.
The size can be stated in terms of $\rho(1/e)$, the value of $\rho$
for which the optical density is $1/e$ times the maximum. Since this value is
necessarily deduced from a two-dimensional image of the cloud and thus
includes the effects of all atoms in the line of sight, it is not exactly
equivalent to the $\rho(1/e)$ parameter extracted from calculations as above.
Nevertheless, we examine in Fig. 4 how BEC is manifest with data on
$N$ and $\rho(1/e)$, as defined above, for various assumed temperatures.
These computational results are for $T>T_{c}$ and were obtained with
$a$=0. Over the range of interest, the $N$ vs. $\rho(1/e)$ curves are quite
flat due to the condensation process shown in Figs. 2 and 3.  To characterize
the quantum state distribution, we have plotted a few values of $N_{0}/N$ as an
orthogonal grid.  The data point reported in Ref. \cite{Hulet} is clearly
within the condensation regime.

An alternative discussion of the observations on $^7$Li has been developed in
terms of rotation of the cloud \cite{Stringari}.  It has been found (and
we confirm) that if all atoms have, for example, 3 quanta of angular momentum
around the symmetry axis, about 8,000 atoms in the ground state will be stable. We find further
that if all atoms have $J_{z} = 30$, a condensate of about 50,000
atoms will be stable.  However if the total value of $J_{z}$ were not an
integer times $N$, there will be differing rates of rotation, hence a
possibility for collisions, in which case the distribution over $m_{i}$ would
be statistical.  To consider such a distribution, we have added an angular
velocity term to the Bose distribution law, Eq. (3), giving an exponent
$[(\epsilon_{i} - \epsilon_{0})\hbar\omega - \mu - \gamma m_{i}]/kT$,
where $m_{i}$ is the value of $J_{z}$ for the $i$th level and $\gamma$
is an additional Lagrange multiplier analogous to that used for fermions in
the ``cranking model'' of nuclear physics \cite{BohrM}. With this
distribution, one does not obtain a concentration of atoms at a single high
value of $m_{i}$.  It seems possible that the cloud of atoms in the trap
may randomly acquire angular momentum when loaded, but this is not needed
to explain the observations.

In conclusion, one may debate the semantic question as to what consitutes BEC,
but self-consistent mean field results presented here for $^7$Li atoms show
that a ``condensate peak'' in the spatial distribution developes rapidly for
$T$ near $T_{c}$. The experimental data \cite{Hulet,RHulet} on $N$ vs.
$\rho(1/e)$ span the region of rapid decreasing
$\rho(1/e)$. It is predicted that over a small range of $N$ for a given $T$, a
condensate peak does occur, but for $N$ slightly greater or $T$ slightly less,
the cloud will collapse.  Confirmation of this and other predictions with
experimental data would be useful tests of $T>0$ mean field theory for bosons.

The author is indebted to R. Hulet, C. Bradley and C. Sackett for many
communications regarding the experiments on $^7$Li, and for helpful
discussions with E. Shuryak, C. N. Yang, N. Balazs, J. Marburger, M. Edwards,
H. Metcalf and P. Koch.  This work was supported by the NSF, ONR, and by a
grant of computer time from the Cornell Theory Center.

\begin{figure}
\label{fig1}
\caption{Calculated distribution of $^7$Li atoms in a harmonic trap as at Rice
University, with $T$ = 150 nK.  The plotted curves show
distributions for several values of the chemical potential, $\mu$, which
determines the atom number, $N$. The light lines are for an ideal gas with
$a$=0, while the heavy lines pertain to $a$=-27.2 $a_{0}$, as for $^7$Li.  The
curve with smallest $N$ is identical to that for a Boltzmann distribution.}
\end{figure}

\begin{figure}
\label{fig2}
\caption{Calculated spatial distribution of 20,000 $^7$Li atoms in a trap
as in Fig. 1,  with $a=-27.2 a_{0}$. The ``condensate peak'' arises over a
small range of $T$. The peak includes atoms in the several lowest quantum
states, not simply the ground state.}
\end{figure}

\begin{figure}
\label{fig3}
\caption{Condensation phenomena near $T=T_{c}$, for 20,000 $^7$Li atoms
in a trap as in Fig. 1.
The solid line gives the $\rho$ value at $1/e$ times the maximum density
(whether the distribution is Gaussian or not) as a function of intensity
for $a$=0 (ideal gas), while the circles give the same parameter calculated
with $a=-27.2 a_{0}$, taken from the distributions shown in Fig. 2. (Both
refer to the outer left axis.) The long dashed line gives $N_{0}/N$ on a
logarithmic scale (right axis).  The short dashed line and inner left axis
show $\langle E \rangle/3k$ from averaging over all quantum states in the
distribution.}
\end{figure}

\begin{figure}
\label{fig4}
\caption{Computational results obtained with $a$=0, plotted in the form of
parameters measured experimentally, namely the total number of atoms $N$ and
the $1/e$ width of the distribution in the $\rho$ direction.  The roughly
horizontal lines are for constant $T$ as shown at the left, while the
approximately vertical lines indicate various values of $N_{0}/N$.  A data
point from Ref. 2 is shown.  The rapid variation of $\rho(1/e)$ with
$N$ for a given $T$ reflects the occurrence of Bose-Einstein condensation.}
\end{figure}

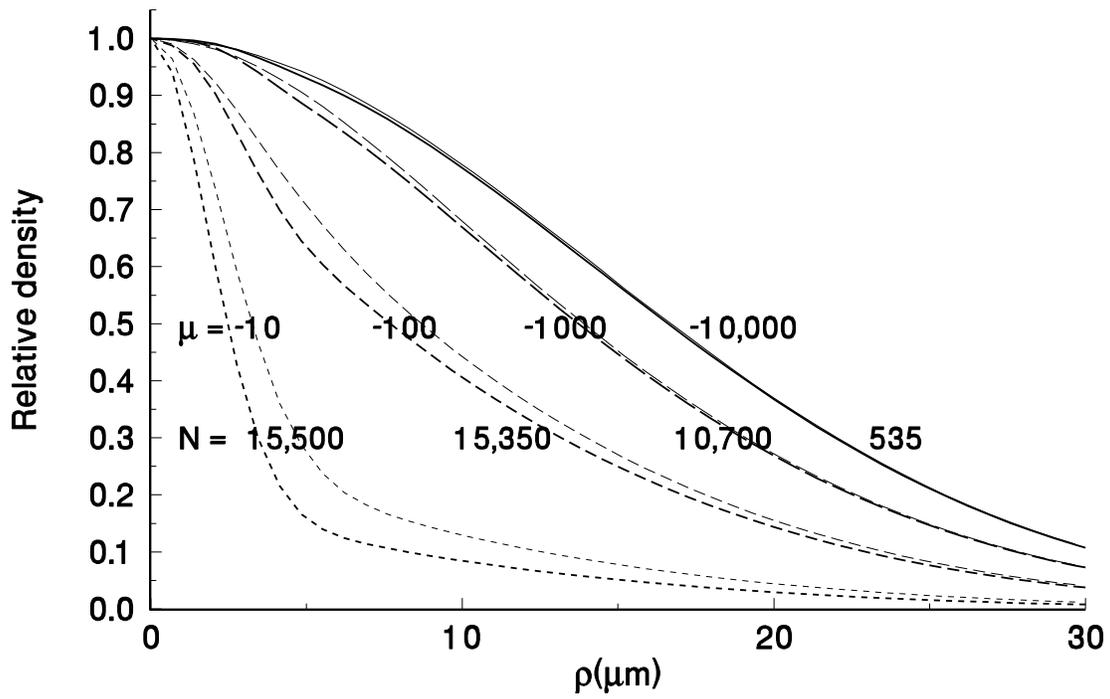
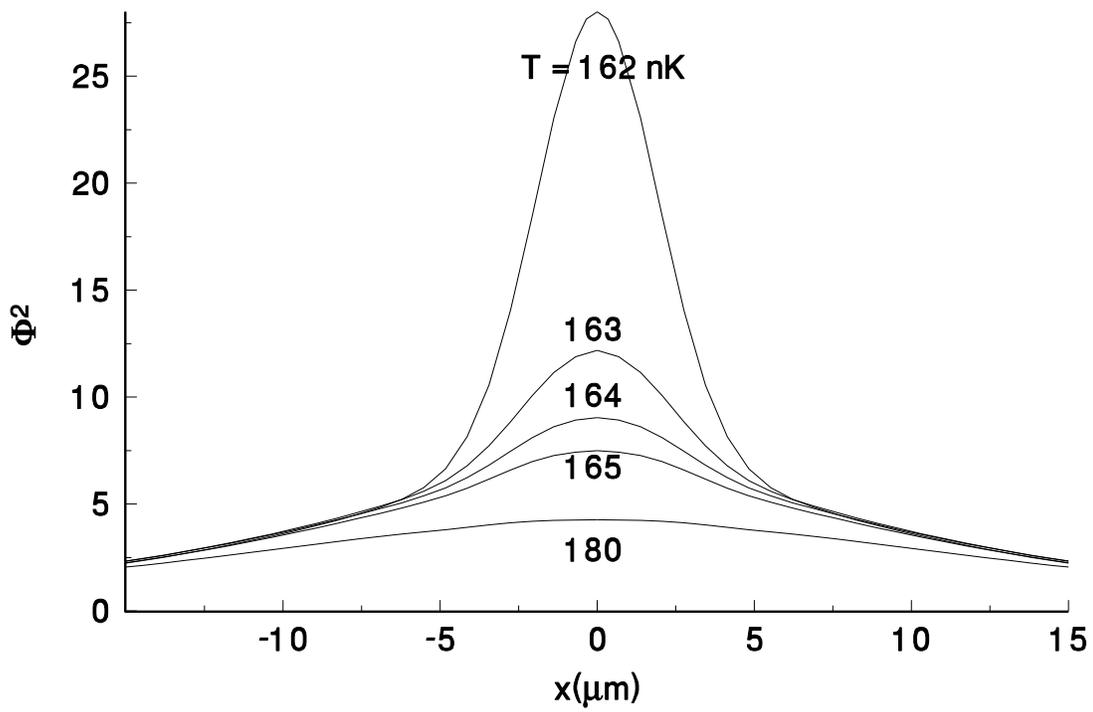
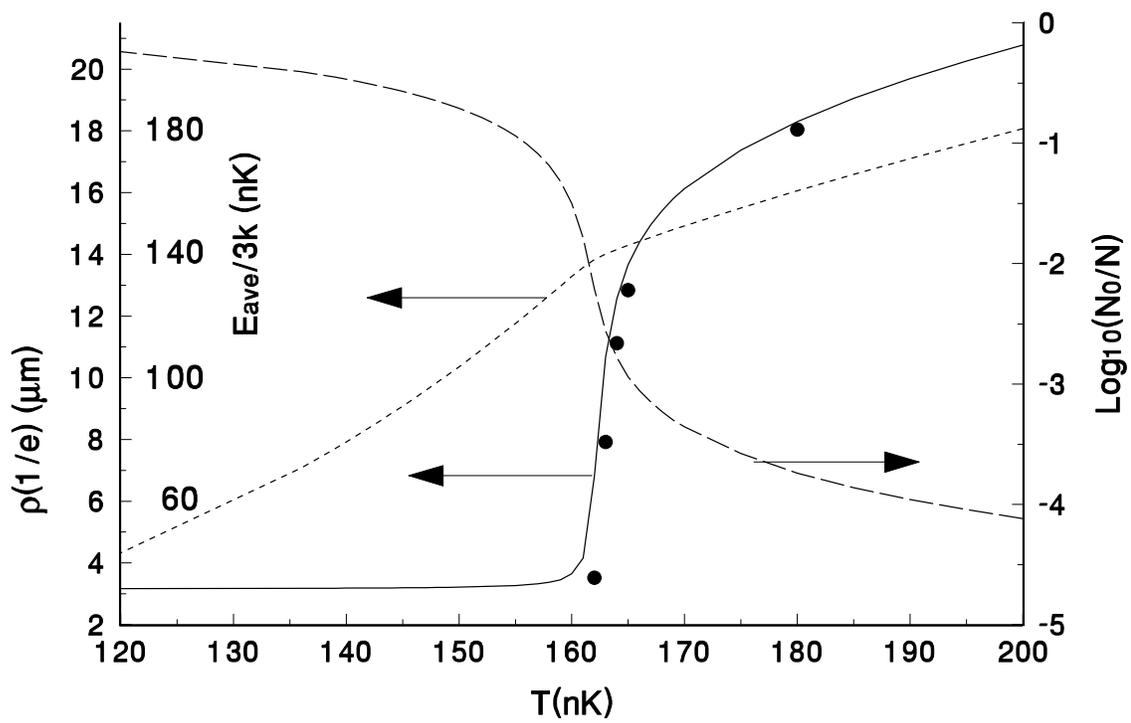
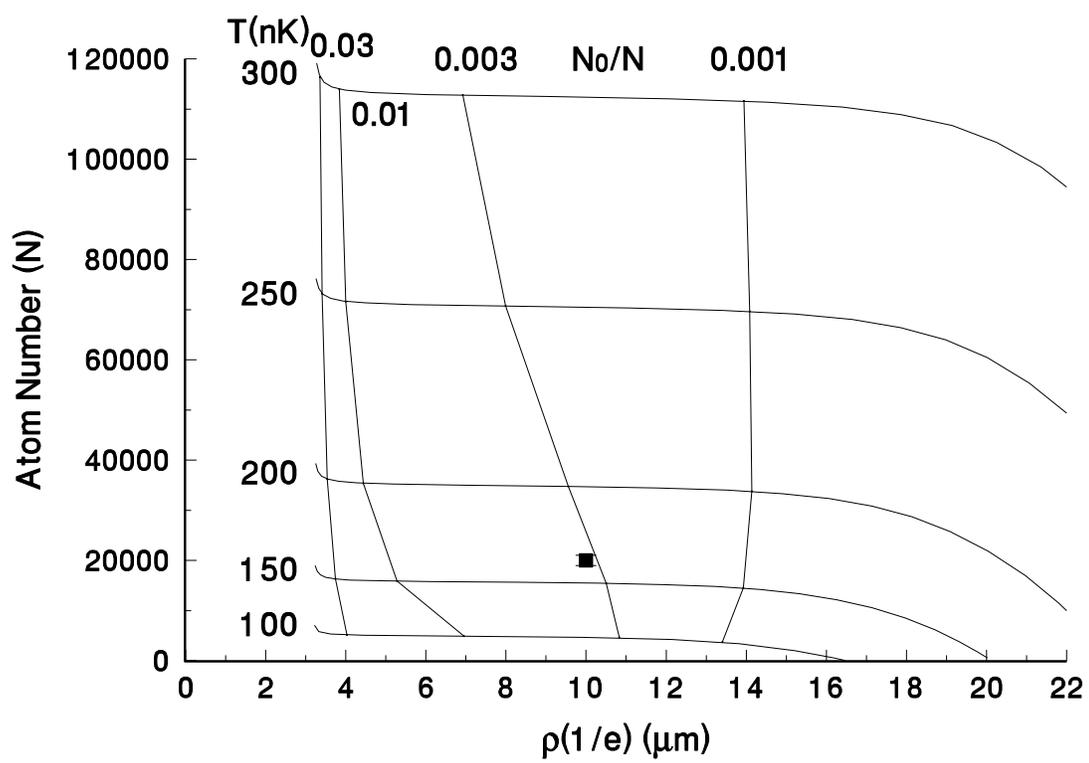
\end{document}